\documentclass{article}
\usepackage{spconf,amsmath,graphicx}
\title{Content-based Representations of audio using Siamese neural networks}
\name{Pranay Manocha $^{\dagger \diamond}$, Rohan Badlani $^{\star \diamond}$, \thanks{$^{\diamond}$ First two authors contributed 
equally} Anurag Kumar $^{\ddag}$, Ankit Shah $^{\ddag}$, Benjamin Elizalde $^{\ddag}$$^{\mathsection}$\thanks{$^{\mathsection}$Acknowledges CONACYT for his doctoral fellowship, No.343964}, Bhiksha Raj $^{\ddag}$  }

\address{$^{\dagger}$Department of Electronics and Electrical Engineering IIT Guwahati, India\\
 $^{\star}$Department of Computer Science, BITS Pilani, India \\
$^{\ddag}$Language Technologies Institute, Carnegie Mellon University, Pittsburgh, United States\vspace{2mm}\\
Email: pranaymnch@gmail.com, rohan.badlani@gmail.com, alnu@andrew.cmu.edu, \\ aps1@andrew.cmu.edu, bmartin1@andrew.cmu.edu, bhiksha@cs.cmu.edu
}

\begin{document}
\ninept
\maketitle
%
%\vspace{-0.1in}
\begin{abstract}
In this paper, we focus on the problem of content-based retrieval for audio, which aims to retrieve all semantically similar audio recordings for a given audio clip query. This problem is similar to the problem of query by example of audio, which aims to retrieve media samples from a database, which are similar to the user-provided example.  We propose a novel approach which encodes the audio into a vector representation using Siamese Neural Networks. The goal is to obtain an encoding similar for files belonging to the same audio class, thus allowing retrieval of semantically similar audio. Using simple similarity measures such as those based on simple euclidean distance and cosine similarity we show that these representations can be very effectively used for retrieving recordings similar in audio content.

\end{abstract}
\vspace{-0.1in}
\section{Introduction}
\label{sec:intro}
Humans have an inherent ability to distinguish and recognize different sounds. Moreover, we are also able to relate and match similar sounds. In fact, we have the capability to detect and relate sound events or ``acoustic objects'' which we have never encountered before, based on how that phenomenon stands out against the background \cite{kumar2014detecting}. This ability plays a crucial role in our interactions with the surroundings and it is also expected that machines have this ability to relate the two audio recordings based on their semantic content. This is precisely the broad goal of this paper. We propose a method to encode semantic content of an audio recording such that two audio recordings of similar content (containing same audio events) can be matched and related through these embeddings. More specifically, we address the problem of content-based retrieval for audio: given an input audio recording we intend to retrieve audio recordings which are semantically related to it. 

Semantic similarity matching and retrieval based on it has received much attention for video and images \cite{WangSiameseImageText}, \cite{OngSiameseImage} and \cite{QiSketchBasedImage}. However, in the broad field of machine learning for audio, semantic similarity matching and retrieval based on audio has received limited attention~\cite{wold1996content}. A major focus has been music information retrieval \cite{CaseyMusicRetrieval}, \cite{Lew:2006:CMI:1126004.1126005}, \cite{foote1997content} and semantic retrieval of audio using text queries \cite{Chechik:2008:LCA:1460096.1460115}, \cite{Patil2016TextRetrieval}. Our focus here is on non-music and non-speech content, sounds which we hear everyday in our daily life, since they play an important role in defining the overall semantic content of an audio recording. Note that the problem of semantic content matching is different from the problem of audio event detection and classification \cite{DCASEPaper}. Our goal is not to detect or classify sound events in an audio recording but to develop methods which capture the semantic content of an audio and be useful in retrieving similar audios. One method which has been explored considerably for audios is the idea of fingerprinting. 
%In \cite{zhangRetrieval}, songs are acoustically represented as HMM’s trained on low-level sound features, with similarity defined as the closeness to each song model. Other approaches like \cite{Vignoli:ISMIR05} consider audio clips as a probability distribution of timbre feature vectors, and similarity is based on Kullback-Leibler divergence.

Audio fingerprinting is an acoustic approach that provides the ability to derive a compact representation which can be efficiently matched against other audio clips to compare their similarity or dissimilarity \cite{DanEllisAudFingerPrint}. Audio fingerprinting has various applications like Broadcast Monitoring\cite{haitsma2002highly}, Audio/Song Detection\cite{wang2003industrial}, Filtering Technology for File Sharing\cite{sherlock1994fingerprint} and Automatic Music Library organization\cite{cano2005content}. 
% Fingerprinting has been used in various subfields of music recognition like \cite{cotton2010audio}, \cite{wang2003industrial}, \cite{ogle2007fingerprinting}, \cite{shrstha2007synchronization} and \cite{bryanclustering}. 

%and mostly requires retrieval of clips that contain similar audio events. For example, given a query of a high-pitched chirping bird clip, a pure fingerprinting based system might retrieve clips which have exactly the same bird chirping, however many times the intent is to retrieve sounds of different birds.

%Many nearest neighbor clustering approaches like \cite{Pampalk05improvementsof} use spectral features which indicate that purely acoustic based approaches may not yield best performance and that a semantic level understanding of the audio content is required to develop a state of the art audio retrieval system. 
We focus on developing an efficient content-based retrieval system that can retrieve audio clips which contain similar audio events as the query audio clip. One can potentially think of applying the conventional fingerprinting approach  \cite{wang2003industrial} for matching  to find recordings with similar audio content. However, fingerprinting is useful only in finding \emph{exact match}. It has been used for finding multiple videos of the same event \cite{cotton2010audio}. In \cite{ogle2007fingerprinting} it is used to find multiple occurrences of a sound event in a recording. But it cannot solve the problem of retrieving all semantically similar files together. In fact even for finding repetitions of the same sound event it does not work well if the sound event is unstructured \cite{ogle2007fingerprinting}. The reason is that fingerprinting tries to capture local features specific to an audio recording. It does not try to capture the broader features which might represent a semantically meaningful class.  For searching similar videos based on content, we need audio retrievals belonging to the correct audio class and not just exact matches as in conventional fingerprinting. Hence, we need representations which can encode class specific information. In this paper, we try to achieve this by using a Siamese Neural Network. Although the siamese network has been previously explored for representations and content understanding in images and video \cite{wang2015unsupervised, OngSiameseImage}, to the best of our knowledge this is the first work employing it in the context of sound events.

Siamese neural networks incorporate methods that excel at detecting similar instances but fail to offer robust solutions that may be applied to other types of problems like classification. In this paper, we present a novel approach that uses a Siamese network to automatically acquire features which enable the model to distinguish between clips containing distinct audio events and encodes a given audio into a vector fingerprint. We show that the output feature vector has an inherent property to capture semantic similarity between audio containing same events. Although the cost of the learning algorithm itself may be considerable, this compressed representation is powerful as we are able to not only learn them without imposing strong priors like in \cite{wang2003industrial}, but also to retrieve semantically similar clips by using this feature space.

\section{Proposed Approach}
\label{sec:format}
\vspace{-0.1in}
\begin{figure}[h]
\centering
  \includegraphics[width=0.5\textwidth]{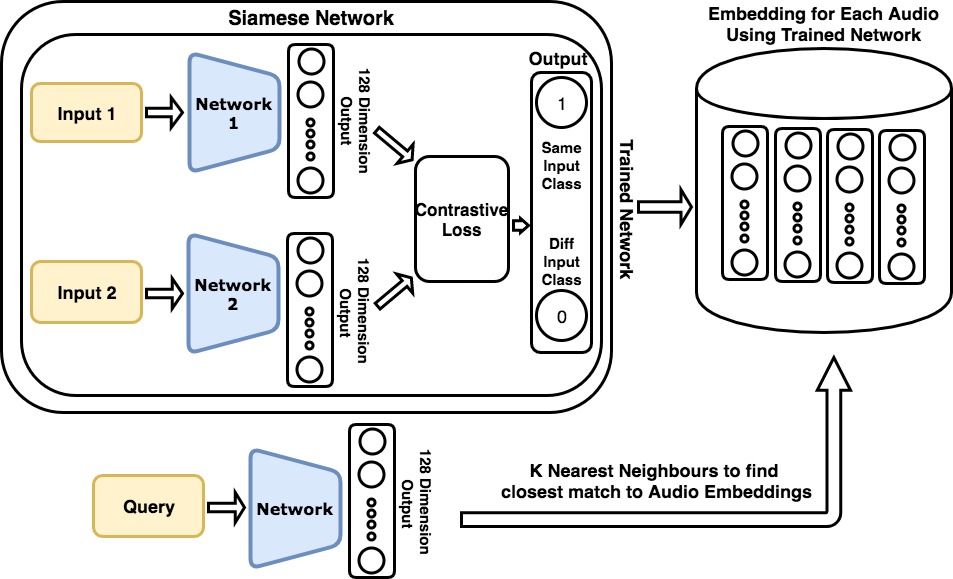}
  \caption{Framework of the Proposed approach}
  \vspace{-0.2in}
  \label{fig:proposed_approach}
\end{figure}
\subsection{Framework Outline}
We propose a neural network based approach to obtain representations or embeddings such that semantically similar audios have similar embeddings. 
%for the problem of similarity matching where we use a Siamese Network trained using a unique and novel contrastive loss function borrowed from \cite{hadsell2006dimensionality}. This leads to fast training and results in better retrieval precision on top K retrieved audio classes for a query.
 
%We learn semantic representations for audio clips via a supervised metric-based approach with Siamese neural networks, then reuse that network’s features for one-shot learning without any retraining. The Siamese twin network returns a feature vector for each audio clip. As illustrated in Fig \ref{fig:tsne_plot} using the TSNE plot \cite{van2013barnes}, clips belonging to the same audio event cluster together in this feature space which show that this feature representation captures distinctions between distinct audio events. We obtain the feature vector for the query audio clip using the network and we return the k nearest neighbors in this feature space, as a ranked list. Fig \ref{fig:proposed_approach}  shows our proposed framework for content based audio retrieval.

We learn these semantic representations through Siamese Neural Networks. Fig \ref{fig:proposed_approach} shows our proposed framework. A Siamese neural network actually consists of two twin networks. The Siamese network takes in two different inputs, one applied to each twin, and is trained to learn the similarity between these inputs. If the inputs are similar then it should predict $1$ otherwise $0$. We use the trained component (twin) network as a feature extractor to obtain representations for audio recordings in the database, as shown in the figure. The input audio query is also embedded through the same network and its embedding is matched with embeddings of recordings in the database to rank them in decreasing order of similarity. This ranking can be done through any distance or similarity measure. In this work we use cosine similarity and euclidean distance. Based on the ranked list one can return the \emph{top K} most similar audios.   

%\begin{figure}[h!]
%\centering
%\includegraphics[width=0.5\textwidth]{1}
%  \caption{TSNE plot for 3 classes- Clock Tick Sound, Brushing Teeth Sound and Pig Sound}
%  \label{fig:tsne_plot}
%  \vspace{-0.1in}
%\end{figure}

\vspace{-0.1in}
\subsection{Siamese Network and Loss Function}
\vspace{-0.1in}
\label{ssec:subhead13}
The Siamese neural network is a class of neural network architectures that contains two or more identical sub-networks, meaning that all sub-networks have the same configuration with the same parameters. Weights and the parameter updates are mirrored across all sub-networks simultaneously. Siamese networks have previously been used in tasks involving similarity or identifying relationships between two or more comparable things. Muller et al.\cite{mueller2016siamese} used a Siamese network for paraphrase scoring by giving a score to a pair of input sentences. Bromley et al.\cite{bromley1994signature} used a Siamese network for the task of signature verification. In the domain of audio, it has been incorporated for content-based matching in music \cite{raffel2015large} and in speech to model speaker related information \cite{zeghidour2016joint, chen2011extracting}. 

Siamese networks offers several advantages. All subnetworks have similar weights which leads to fewer training parameters thus requiring less training data and a lesser tendency to over fit. Moreover, the outputs of each of the subnetworks are representation vectors with the same semantics and this makes them much easier to compare with one other. These characteristics makes them well suited for our task. 

To train our Siamese Network we use the contrastive loss function defined in \cite{hadsell2006dimensionality}. The goal is to learn the parameters W of a function $G_{W}$, such that neighbors are pulled together and non-neighbors are pushed apart. To achieve this, the objective function needs to be suitably defined. The learning process here operates on a pair of samples. Let {$X_1$ , $X_2$} $\in$ $\mathcal{P}$ be a pair of input samples and let $Y$ be the label assigned to this pair. $Y=1$ if the inputs $X_1$ and $X_2$ are similar, otherwise $Y=0$. The distance between $X_1$ and $X_2$ is defined as the euclidean distance between the mapping from the function $G_{W}$
\begin{equation}
\label{eq:dist}
D_{W}(X_1, X_2) = \parallel G_{w}(X_1)-G_{w}(X_2)\parallel
\end{equation}

The overall loss function for the pair $X_1$ and $X_2$, is then defined as
\begin{align}
\label{eq:lossfn}
\resizebox{0.9\columnwidth}{!}{
$L(W, Y, X_1, X_2) = (Y)\frac{1}{2}(D_{w})^2 + (1-Y)\frac{1}{2}\{max(0,m-D_{w})\}^2 $
}
\end{align}
In Eq \ref{eq:lossfn},  $m > 0$ is the margin. The idea behind this margin is that the dissimilar points contribute to the training loss only if the distance between them, $D_{W}$, is within the radius defined by margin value $m$. For the pairs of similar inputs we always want reduce the distance between them. 

\begin{figure}[t!]
\centering
  \includegraphics[width=0.9\columnwidth]{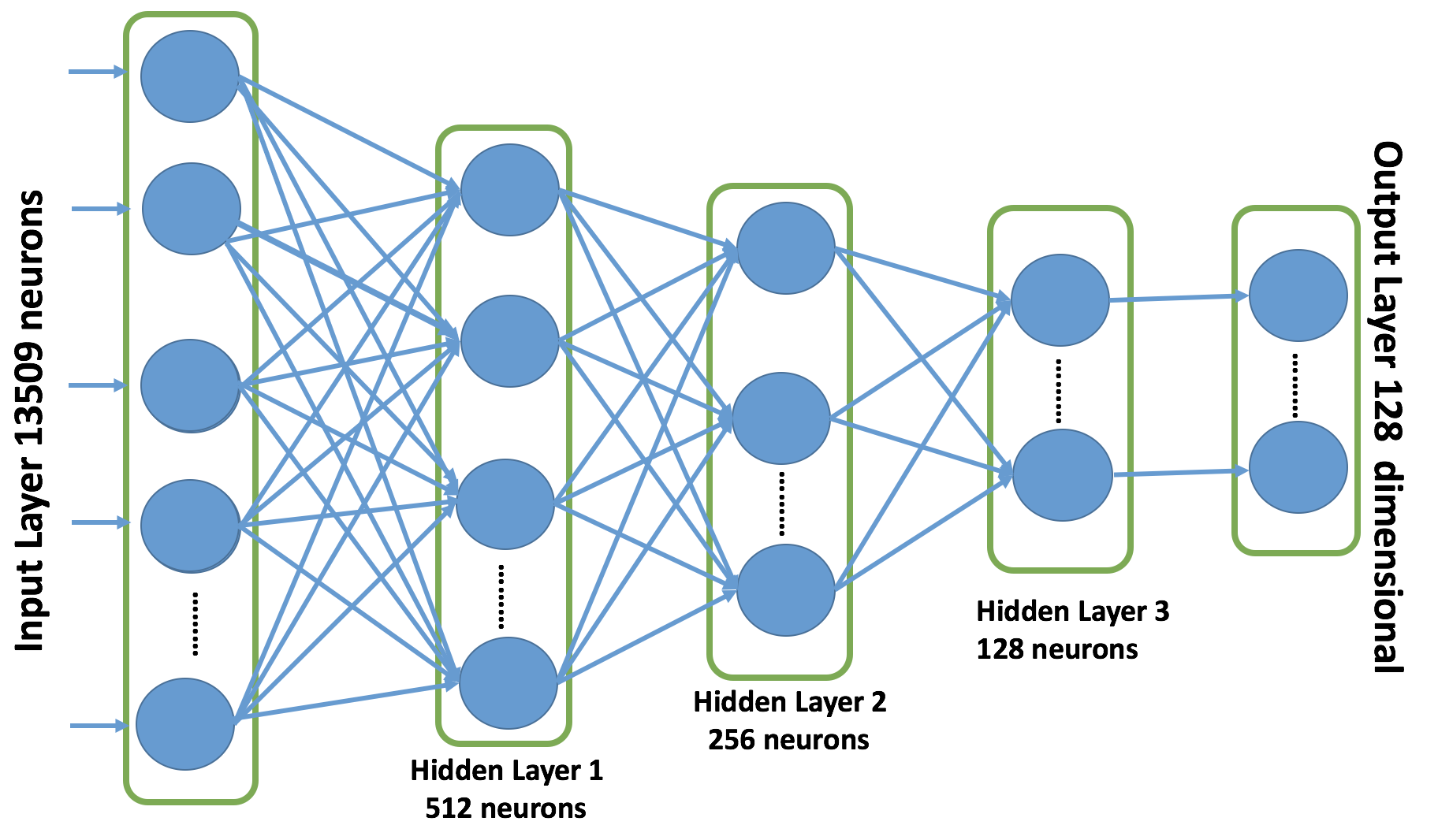}
  \caption{Architecture of the Subnetworks in the Siamese Network. The final layer of $128$ neurons is also the output layer}
  \label{fig:architecture}
  \vspace{-0.25in}
\end{figure}
\vspace{-0.1in}
\subsection{Network Architecture}
\vspace{-0.1in}
\label{ssec:subhead10}
The architecture of the individual sub-networks in the Siamese network is shown in Fig \ref{fig:architecture}. Each sub-network is a feed-forward multi layer perceptron (MLP) network. The input to the network are log-frequency spectrograms  of audio recordings. The frames in Logspec are concatenated to create one long extended vector. The dimensionality of the inputs are $13509$ (See \ref{ssec:strain} for details). The network consists of a total of $3$ layers after the input layer. The first layer consists of $512$ neurons, the second layer $256$ neurons and the last layer has $128$ neurons. The last layer also serves as the output layer. The activation function in all layers is ReLU ($max(0,x)$). A dropout of $0.3$ is applied between all three layers during training. We will refer to the network as $\mathcal{N}_R$

\vspace{-0.1in}
\subsection{Representations and Retrieval}
\vspace{-0.1in}
All audio clips in the audio database are represented through the $128$ dimensional output from the network $\mathcal{N}_R$. When a query audio clip is given, we first obtain its $128$ dimensional representation using $\mathcal{N}_R$. This representation is then matched to representations of all audios in the database using a similarity or distance function. The  clips in database are ranked according to the similarity measure and then the top K clips are returned. In other ways, one can think of it as obtaining $K$ nearest neighbors in the database. Note that all operations are done on fixed length audios of $2$ seconds, details are provided in further sections. 
%Variable length training audio clips and retrieval using variable length query are actively being investigated and will be part of the future work. 

\vspace{-0.2in}
\section{DATASET AND EXPERIMENTAL SETUP}
\vspace{-0.1in}
We study the problem of semantic similarity based retrieval in the domain of sound events. More specifically, given an audio clip of a sound class, the goal is to retrieve all audio clips of that class from the database. We consider the list of sound events from $3$ databases, ESC-50\cite{piczak2015environmental}, US8K\cite{Salamon:UrbanSound:ACMMM:14} and TUT 2016\cite{Mesaros2016_EUSIPCO}. Overall, we considered a total of $76$ sound events. Some examples are, \emph{Dog Barking, Clock Tick-Tock, Wind Blowing } etc. audio  events include wide range of sound events, including different sound events from broad categories from animal sounds such as \emph{Dog Barking and Crow}, non-speech human sounds such as \emph{Clapping and Coughing}, exterior sounds such as \emph{Siren, Engine, Airplane} to urban soundscape sounds such as \emph{Street Music, Jackhammer} etc. 

\vspace{-0.1in}
\subsection{YouTube Dataset}
\vspace{-0.1in}
\label{ssec:subhead11}
The importance of semantic similarity matching lies in its utility in content-based retrieval of multimedia data on the web. Hence, we work with audio recordings from YouTube. User generated recordings on multimedia sharing websites such as YouTube, are often very noisy, unstructured and recorded under different conditions. Hence, even \emph{intra-class} variation is very high, which makes content-based retrieval an extremely difficult problem. 

For each of the $76$ classes, we obtain $100$ recordings from YouTube. To obtain relevant results we use \textless SOUND\_NAME sound \textgreater  ( e.g \textless car horn sound \textgreater) as search query. From the returned list we select $100$ recordings based on their length and  relevance for the query. Very short (\textless 2 seconds) and very long (\textgreater 10 min) recordings are not considered.

%The dataset consists of 10 batches of 760 files each, each batch having 76 classes. We use the audio events from three different datasets - ESC50\cite{piczak2015environmental}, US8K\cite{Salamon:UrbanSound:ACMMM:14} and TUT 2016\cite{Mesaros2016_EUSIPCO} containing 50, 10 and 18 audio events respectively. The main idea behind creating an aggregated audio event list was to ensure scalability over audio events in the future and ensuring that the model is able generalize to audio events from different datasets. 

%Since the audio query to YouTube with just the audio event name does not yield good results, we append the suffix "sound" to the audio event, for example, the corresponding query for audio event 'car horn' is 'car horn sound'. After appending these labels with suitable suffixes, these were queried to YouTube and the top instances were crawled using Pafy python module and indexed by their youtube Ids. Videos longer than 10 minutes and shorter than 3 seconds were disregarded because either they were unrelated files or too long/short to be processed. 

We divided the dataset in the ratio of 70-10-20. 70 percent of the data per class is used for training and the remaining 30 percent data is split 1:2 between validation and testing. Thus, we take 70 samples per class for training, 10 for validation and the remaining 20 for testing, given that we roughly have 100 files per audio class. Overall, we have around 5K audio files for training, 760 files for validation and around 1500 files for testing. For our experiments, we operate on $2$ second clips from each of these recordings. Hence, our actual database for experiments are fixed length $2$ second audio clips in training as well as validation and test sets.  

\vspace{-0.1in}
\subsection{Siamese Network Training}
\vspace{-0.1in}
\label{ssec:strain}
The inputs to the Siamese network must be pairs of audio clips. We assign label 1 to the pairs of clips from the same class and label 0 to the pairs from different classes. We consider two training sets, balanced and unbalanced. The network trained on balanced set will be referred to as $\mathcal{N}_R^B$ and that trained on unbalanced as $\mathcal{N}_R^{U}$. In the balanced case, to create pairs with positive label ($Y=1$), we consider all possible pairs belonging to the same audio class. For pairs with negative label ($Y=0$), a clip belonging to a sound class is randomly paired with a clip from any other sound class. Hence, we end up with equal number of positive and negative label pairs. In the unbalanced case the positive label pairs are obtained in the same way. But for the negative label, we pair a clip belonging to a sound class with all clips not belonging to that sound. Thus, we have a non equal distribution of positive and negative labels.

We used the log spectrogram features, taking 1024 point FFT over a window size of 64ms and an overlap of 32ms per window. Both the axis were converted to the log scale and 79 bins were chosen for the frequency axis whereas 171 quantization bins were chosen for the time scale. We then concatenate these 79X171=13509 features and use as an input to the Siamese Network. The reason for taking log spectrogram features is that the features having a large difference on a common scale have diminished differences on the log scale. It is specially useful in those cases where we have a huge variation of feature magnitudes and hence it brings all a common scale. All parameters were tuned using the validation set. We train each model to 200 epochs and optimize on the training and validation losses. 

\vspace{-0.2in}
\subsection{Retrieval}
\vspace{-0.1in}
\label{ssec:retrive}
We obtain the vector encoding of each file of the audio class by passing it through the trained saved model. All audio clips in the database are represented by these representations. At the time of testing, we obtain the representation for the query audio clip using the network and compute its similarity with representations of audios in the database. For computing similarity between two representations we use either euclidean distance or cosine similarity. 
\vspace{-0.15in}
\begin{figure}[h!]
%\centering
  \includegraphics[width=0.8\columnwidth]{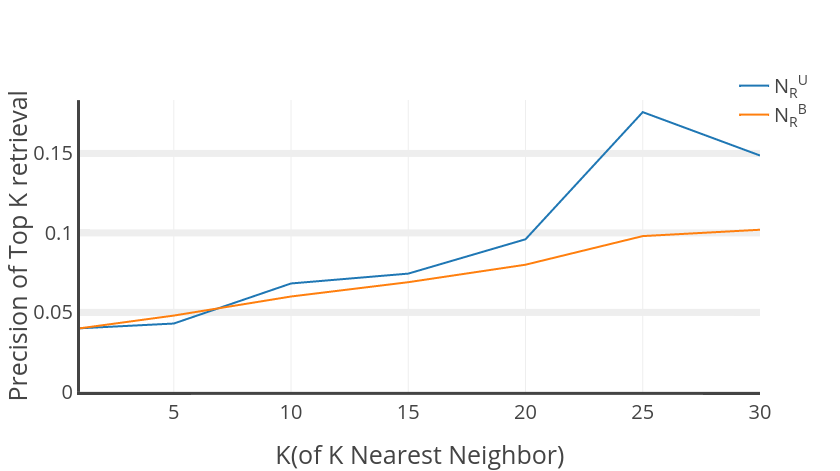}
  \caption{Variation of $MP^{K}$ with different K, from K=1 to K=30}
  \vspace{-0.2in}
  \label{fig:kselection}
\end{figure}

\vspace{-0.15in}
\section{Evaluation and results}
\label{sec:experiments}
\vspace{-0.1in}
\subsection{Metrics}
\vspace{-0.1in}
For any given query audio, we obtain a ranked list of audio clips present in the database which contain similar audio events present in the query clip. We then compute 3 metrics for evaluation which are defined below:
\vspace{-0.1in}
\subsubsection{Average Precision}
\vspace{-0.1in}
\label{ssec:ap}
The average precision for a query is defined as mean of precisions at all positive hits. 
\begin{equation}
AP=\frac{1}{m_j}\sum_{i=1}^{m_j}Precision_i
\end{equation}
$Precision_i$ measures the fraction of correct items among first $i$ recommendations. This precision is measured at every positive hit in the ranked list. $m_j$ refers to the number of positive items in the database for the query. Average precision is simply the mean of these precision values. We will be reporting the mean of average precision (MAP) over all queries. 

\vspace{-0.1in}
\subsubsection{Precision at 1}
\label{sssec:subsubhead4}
\vspace{-0.1in}
This metric measures the precision at the first positive hit only. The idea is to understand where does the first positive item lie in the ranked list. Again the mean of Precision at 1 ($MP^1$) over all queries are reported. 

\vspace{-0.1in}
\subsubsection{Precision of Top K retrieval}
\label{sssec:subsubhead6}
\vspace{-0.1in}
This metric measures the quality of retrieved items in the top $K$ items in the ranked list. For each query, we calculate the number of correct class instances in the top K files and then divide that by K to get the precision of the correct class amongst the top K retrieved files and take an average across all queries. Multiplying this score by K tells us the average number of correct class matches in the top K of the retrieved list. This measure tells us about the precision of the correct class in the top K retrieved list. Once again the mean of this metric over all queries is reported ($MP^{K}$)

The variation of $MP^{K}$ with $K$ is shown in Figure \ref{fig:kselection}. We observe that this metric is maximum around K=25 and hence we report the best possible performance from now on for K=25.
 
\vspace{-0.2in}
\subsection{Results and Discussion}
\vspace{-0.1in}
\begin{table}[t]
\centering
\resizebox{0.48\columnwidth}{!}{  
  \begin{tabular}{ | c | c | c | c | }
    \hline
    \textbf{Measures} & \textbf{$\mathcal{N}_R^B$}&\textbf{$\mathcal{N}_R^U$}  \\ \hline
MAP  & 0.0241 & 0.0342 \\ \hline
$MP^1$ & 0.314 & 0.436  \\ \hline
$MP^{K=25}$  & 0.099 & 0.177 \\ \hline
  \end{tabular} 
  }
  \resizebox{0.48\columnwidth}{!}{  
    \begin{tabular}{ | c | c | c | c | }
    \hline
    \textbf{Measures} & \textbf{$\mathcal{N}_R^B$}&\textbf{$\mathcal{N}_R^U$}  \\ \hline
MAP  & 0.0186 & 0.0133  \\ \hline
$MP^1$  & 0.132 & 0.333  \\ \hline
$MP^{K=25}$  & 0.105 & 0.133  \\ \hline
  \end{tabular} 
  }
\caption{Left: Performance using euclidean distance, Right: Performance using cosine similarity}
\label{performance_table}
\vspace{-0.1in}
\end{table}

\begin{figure}[t]
\centering
  \includegraphics[width=\columnwidth]{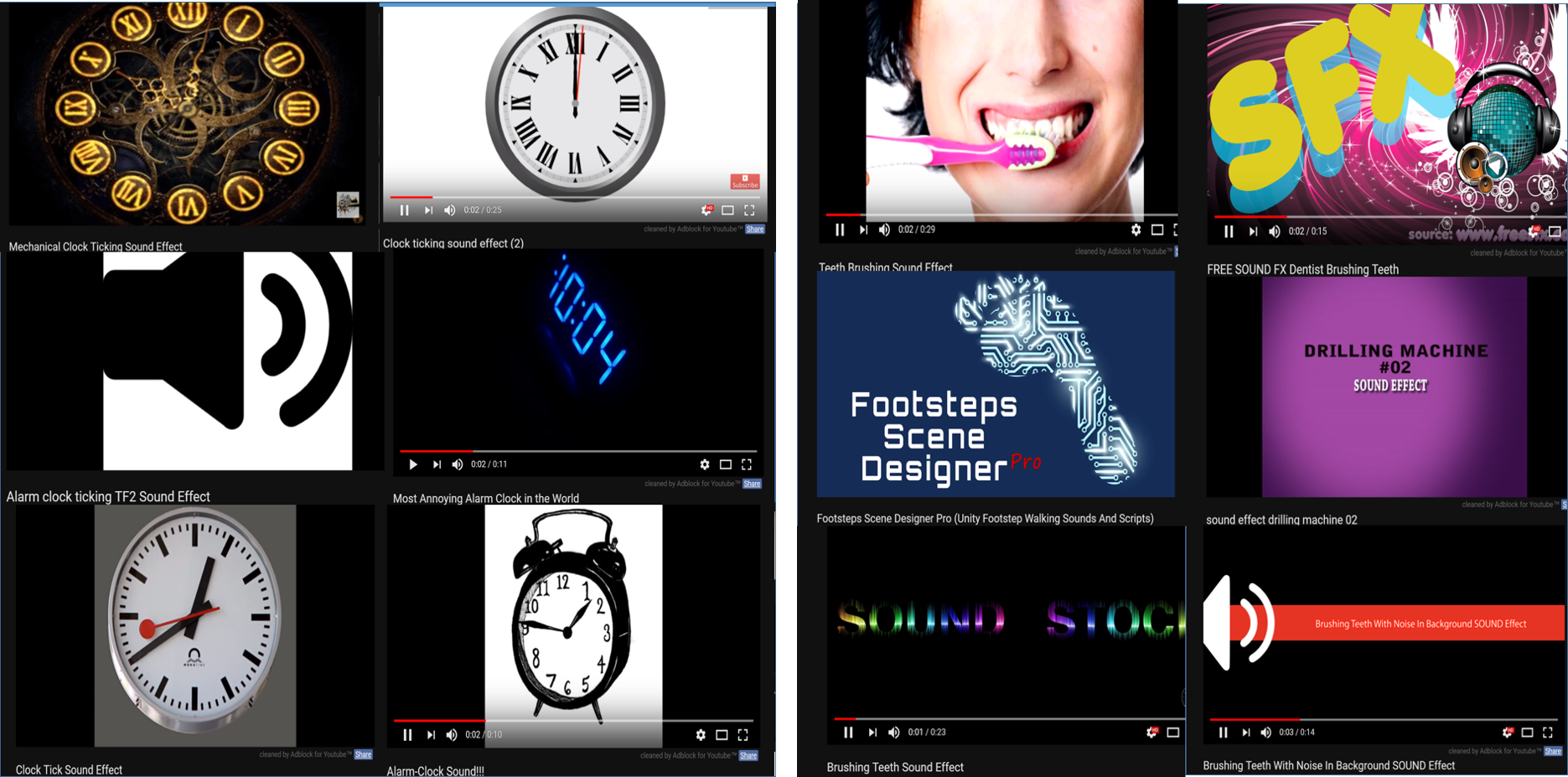}
  \caption{Examples of content-based retrieval. Left: Clock Tick, Right: Brushing Teeth}
  \label{fig:query_sample_thumbnail}
  \vspace{-0.25in}
\end{figure}

% Top 3 MAP classes

\begin{table}[t]
%\centering
\resizebox{0.40\columnwidth}{!}{   
  \begin{tabular}{ |  c | c | }
    \hline
    \textbf{Audio Class} & \textbf{}  \\ 
                         & \textbf{MP\textsuperscript{K=25}}  \\ \hline
Wind Blowing	        &        0.784  \\ \hline
Sheep	        &        0.753  \\ \hline
Pig	& 0.724 \\ \hline
Water Drop	& 0.711 \\ \hline
Clock Tick	& 0.708 \\ \hline
Brushing Teeth	& 0.699 \\ \hline
Drilling Sound	& 0.681 \\ \hline
Helicopter	& 0.670 \\ \hline
Chirping Birds	& 0.650 \\ \hline
Rooster	& 0.636 \\ \hline

  \end{tabular} 
}
\resizebox{0.55\columnwidth}{!}{   
  \begin{tabular}{ |  c | c | }
    \hline
    \textbf{Audio Class} & \textbf{}  \\ 
                         & \textbf{MP\textsuperscript{K=25}}  \\ \hline
Dog Bark	&	 0.110  \\ \hline
Car Horn 	&  	0.272  \\ \hline
Crackling Fire		& 	0.276 \\ \hline
Glass Breaking		& 	0.312 \\ \hline
Can Opening		& 	0.314 \\ \hline
Crying Baby		& 	0.315 \\ \hline
Gun Shot	& 	0.318 \\ \hline
Crickets(insect)	& 	0.41 \\ \hline
Banging Residential Area	& 	0.42 \\ \hline
Children Playing	& 	0.44 \\ \hline
  \end{tabular} 
}
\vspace{-0.1in}
  \caption{Left: Classes (Top 10) with highest precisions, Right:Classes (bottom 10) with least precision}
\label{topandbottomclasses}
\vspace{-0.1in}
\end{table}

We first show performance with respect to queries. From table \ref{performance_table}, we observe that the Euclidean distance performance exceeds the Cosine similarity performance in the Mean Average Precision measure. This may be due to the fact during siamese network training, euclidean distance is used to measure the closeness between two points. Hence, the learned representations are inherently designed to work better with euclidean distance. 

We note that the overall MAP of the system is similar to what has been traditionally observed throughout audio retrieval work \cite{buckley2004retrieval}. $MP^1$ value of around $0.3$ (for $\mathcal{N}_R^B$) indicates that the first positive hit on an average is achieved at rank 3. However, for a given specific query it can be much better.  We note that the $MP^1$ values are fairly high, implying that the first positive hit can be easily obtained using the audio embeddings generated using Siamese Network. Also, note that the network $\mathcal{N}_R^U$ performs much better compared to $\mathcal{N}_R^B$. $\mathcal{N}_R^U$ is trained using a larger set of pairs of dissimilar audios and hence it is able learn more discriminitive representations. 

The most important metric for understanding performance of a retrieval system is $MP^K$. The values for $MP^K$ multiplied by K gives us the average number of correct class instances in the top K for a query. A low value of this measure means that a low number of correct class instances are obtained in the top K retrieved files. We are able to obtain fairly reasonable value of $MP^K$.  

Fig\ref{fig:query_sample_thumbnail} gives visualization of a retrieval example. It shows two examples of queries and their top 6 retrieved similar files. We observe that for the class 'clock tick sound', the retrieval is from classes 'clock tick sound' and 'clock alarm sound', which are both nearly similar audio events. For the class 'brushing teeth sound', the system performs well as their are no other similar audio classes in the database and hence it retrieves 5 out of the 6 files correctly. Overall, it illustrates that our system is capable of delivering content based retrieval of audio recordings. 

We now show performance on retrieval for some specific sound events. We show the average $MP^K$ measure over all queries of a sound event class. Due to space constraints, we are not able to show performance numbers for all classes. In Table \ref{topandbottomclasses}, we show performance for $20$ sound events, $10$ with highest $MP^K$ values and $10$ with lowest $MP^K$ values. First, we note that for several classes we are able to obtain reasonably high performance using our learned representations. For example, for \emph{Wind Blowing}, around $20$ out of the top 25 retrieval actually belong to the sound class wind blowing. However, we also observe that some audio classes have unusually low average precisions. This occurs because we combine the labels of different datasets, there were some similar events classes which are semantically same but were treated as separate classes like 'Dog bark sound' and 'Dog sound' and 'Car horn sound' and 'Car passing by residential area'. Hence, the problem is more of how a sound event is referred to as instead of the actual representation we obtain from our method. We are actively investigating this problem of similar audio events with minor differences in text labels and will be addressing this in our future work.

\begin{figure}[t]
\centering
\includegraphics[width=0.6\columnwidth]{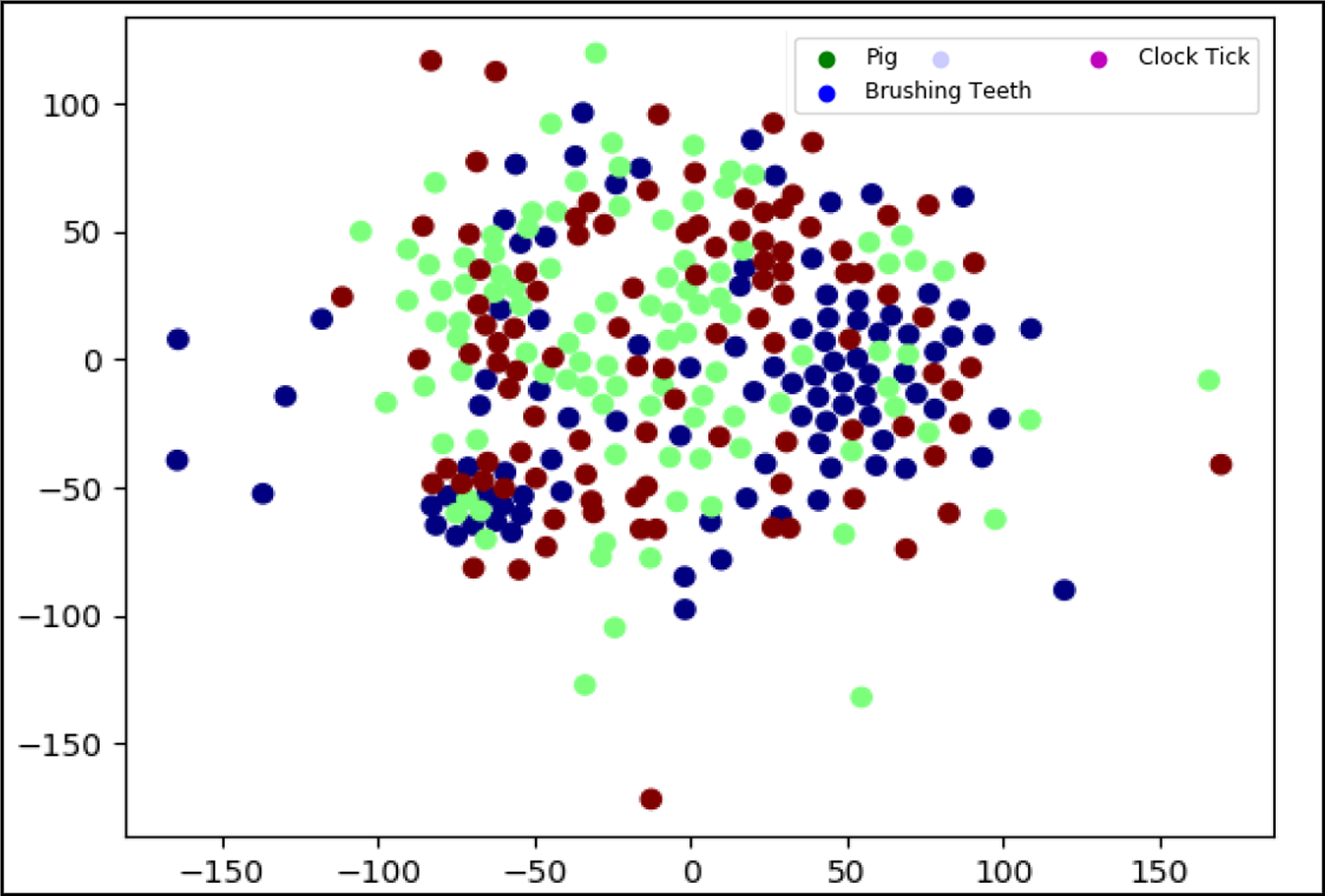}
  \caption{TSNE plot for 3 classes- Clock Tick Sound, Brushing Teeth Sound, Pig Sound}
  \label{fig:tsne_plot}
  \vspace{-0.2in}
\end{figure}

Figure \ref{fig:tsne_plot} shows that the visualization of 2 dimensional t-SNE embeddings for the 3 classes. For the purpose of clarity, we included only 3 classes in the plot. Once can see that the representations learned for audios are actually encoding semantic content as audios of same class cluster close to each other. This demonstrates that the proposed Siamese Network based representations inherently indistinguish between distinct audio events.

\vspace{-0.2in}
\section{Conclusions}
\label{sec:majhead}
\vspace{-0.1in}

We proposed a novel approach that uses Siamese Neural Network to learn representations for audios. Our results indicate that these representations are able to capture semantic similarity between audio containing same events. This makes them well suited for content based retrieval of audio. We observe that for several classes, the precision of top 25 results is very high. We tried different measures of similarity like conventional euclidean distance and Cosine similarity and found that the performance of both of them is similar on retrieval of similar semantic sounds. This shows that the embeddings obtained from the Siamese Neural Network capture the similarity between clips belonging to the audio events very well and can be used for efficient content-based audio retrieval tasks.

\ninept
\bibliographystyle{IEEEbib}
\bibliography{IEEE}

\begin{thebibliography}{10}

\bibitem{kumar2014detecting}
Anurag Kumar, Rita Singh, and Bhiksha Raj,
\newblock ``Detecting sound objects in audio recordings,''
\newblock in {\em Signal Processing Conference (EUSIPCO), 2014 Proceedings of
  the 22nd European}. IEEE, 2014, pp. 905--909.

\bibitem{WangSiameseImageText}
Liwei Wang, Yin Li, and Svetlana Lazebnik,
\newblock ``Learning deep structure-preserving image-text embeddings,''
\newblock pp. 5005--5013, 06 2016.

\bibitem{OngSiameseImage}
Eng-Jon Ong, Syed Husain, and Miroslaw Bober,
\newblock ``Siamese network of deep fisher-vector descriptors for image
  retrieval,''
\newblock 02 2017.

\bibitem{QiSketchBasedImage}
Yonggang Qi, Yi-Zhe Song, Honggang Zhang, and Jun Liu,
\newblock ``Sketch-based image retrieval via siamese convolutional neural
  network,''
\newblock pp. 2460--2464, 09 2016.

\bibitem{wold1996content}
Erling Wold, Thom Blum, Douglas Keislar, and James Wheaten,
\newblock ``Content-based classification, search, and retrieval of audio,''
\newblock {\em IEEE multimedia}, vol. 3, no. 3, pp. 27--36, 1996.

\bibitem{CaseyMusicRetrieval}
M.~A. Casey, R.~Veltkamp, M.~Goto, M.~Leman, C.~Rhodes, and M.~Slaney,
\newblock ``Content-based music information retrieval: Current directions and
  future challenges,''
\newblock {\em Proceedings of the IEEE}, vol. 96, no. 4, pp. 668--696, April
  2008.

\bibitem{Lew:2006:CMI:1126004.1126005}
Michael~S. Lew, Nicu Sebe, Chabane Djeraba, and Ramesh Jain,
\newblock ``Content-based multimedia information retrieval: State of the art
  and challenges,''
\newblock {\em ACM Trans. Multimedia Comput. Commun. Appl.}, vol. 2, no. 1, pp.
  1--19, Feb. 2006.

\bibitem{foote1997content}
Jonathan~T Foote,
\newblock ``Content-based retrieval of music and audio,''
\newblock in {\em Multimedia Storage and Archiving Systems II}. International
  Society for Optics and Photonics, 1997, vol. 3229, pp. 138--148.

\bibitem{Chechik:2008:LCA:1460096.1460115}
Gal Chechik, Eugene Ie, Martin Rehn, Samy Bengio, and Dick Lyon,
\newblock ``Large-scale content-based audio retrieval from text queries,''
\newblock in {\em Proceedings of the 1st ACM International Conference on
  Multimedia Information Retrieval}, New York, NY, USA, 2008, MIR '08, pp.
  105--112, ACM.

\bibitem{Patil2016TextRetrieval}
N.~M. Patil and M.~U. Nemade,
\newblock ``Content-based audio classification and retrieval: A novel
  approach,''
\newblock in {\em 2016 International Conference on Global Trends in Signal
  Processing, Information Computing and Communication (ICGTSPICC)}, Dec 2016,
  pp. 599--606.

\bibitem{DCASEPaper}
D.~Stowell, D.~Giannoulis, E.~Benetos, M.~Lagrange, and M.~D. Plumbley,
\newblock ``Detection and classification of acoustic scenes and events,''
\newblock {\em IEEE Transactions on Multimedia}, vol. 17, pp. 1733--1746, 2015.

\bibitem{DanEllisAudFingerPrint}
D.~Ellis,
\newblock ``Robust landmark-based audio fingerprinting,''
\newblock 09 2009.

\bibitem{haitsma2002highly}
Jaap Haitsma and Ton Kalker,
\newblock ``A highly robust audio fingerprinting system.,''
\newblock in {\em Ismir}, 2002, vol. 2002, pp. 107--115.

\bibitem{wang2003industrial}
Avery Wang et~al.,
\newblock ``An industrial strength audio search algorithm.,''
\newblock in {\em Ismir}. Washington, DC, 2003, vol. 2003, pp. 7--13.

\bibitem{sherlock1994fingerprint}
Barry~G Sherlock, DM~Monro, and K~Millard,
\newblock ``Fingerprint enhancement by directional fourier filtering,''
\newblock {\em IEE Proceedings-Vision, Image and Signal Processing}, vol. 141,
  no. 2, pp. 87--94, 1994.

\bibitem{cano2005content}
Pedro Cano, Markus Koppenberger, and Nicolas Wack,
\newblock ``Content-based music audio recommendation,''
\newblock in {\em Proceedings of the 13th annual ACM international conference
  on Multimedia}. ACM, 2005, pp. 211--212.

\bibitem{cotton2010audio}
Courtenay~V Cotton and Daniel~PW Ellis,
\newblock ``Audio fingerprinting to identify multiple videos of an event,''
\newblock in {\em Acoustics Speech and Signal Processing (ICASSP), 2010 IEEE
  International Conference on}. IEEE, 2010, pp. 2386--2389.

\bibitem{ogle2007fingerprinting}
James~P Ogle and Daniel~PW Ellis,
\newblock ``Fingerprinting to identify repeated sound events in long-duration
  personal audio recordings,''
\newblock in {\em Acoustics, Speech and Signal Processing, 2007. ICASSP 2007.
  IEEE International Conference on}. IEEE, 2007, vol.~1, pp. I--233.

\bibitem{wang2015unsupervised}
Xiaolong Wang and Abhinav Gupta,
\newblock ``Unsupervised learning of visual representations using videos,''
\newblock in {\em Proceedings of the IEEE International Conference on Computer
  Vision}, 2015, pp. 2794--2802.

\bibitem{mueller2016siamese}
Jonas Mueller and Aditya Thyagarajan,
\newblock ``Siamese recurrent architectures for learning sentence
  similarity.,''
\newblock in {\em AAAI}, 2016, pp. 2786--2792.

\bibitem{bromley1994signature}
Jane Bromley, Isabelle Guyon, Yann LeCun, Eduard S{\"a}ckinger, and Roopak
  Shah,
\newblock ``Signature verification using a" siamese" time delay neural
  network,''
\newblock in {\em Advances in Neural Information Processing Systems}, 1994, pp.
  737--744.

\bibitem{raffel2015large}
Colin Raffel and Daniel~PW Ellis,
\newblock ``Large-scale content-based matching of midi and audio files.,''
\newblock in {\em ISMIR}, 2015, pp. 234--240.

\bibitem{zeghidour2016joint}
Neil Zeghidour, Gabriel Synnaeve, Nicolas Usunier, and Emmanuel Dupoux,
\newblock ``Joint learning of speaker and phonetic similarities with siamese
  networks.,''
\newblock in {\em INTERSPEECH}, 2016, pp. 1295--1299.

\bibitem{chen2011extracting}
Ke~Chen and Ahmad Salman,
\newblock ``Extracting speaker-specific information with a regularized siamese
  deep network,''
\newblock in {\em Advances in Neural Information Processing Systems}, 2011, pp.
  298--306.

\bibitem{hadsell2006dimensionality}
Raia Hadsell, Sumit Chopra, and Yann LeCun,
\newblock ``Dimensionality reduction by learning an invariant mapping,''
\newblock in {\em Computer vision and pattern recognition, 2006 IEEE computer
  society conference on}. IEEE, 2006, vol.~2, pp. 1735--1742.

\bibitem{piczak2015environmental}
Karol~J Piczak,
\newblock ``Environmental sound classification with convolutional neural
  networks,''
\newblock in {\em 2015 IEEE 25th International Workshop on Machine Learning for
  Signal Processing (MLSP)}. IEEE, 2015, pp. 1--6.

\bibitem{Salamon:UrbanSound:ACMMM:14}
J.~Salamon, C.~Jacoby, and J.~P. Bello,
\newblock ``A dataset and taxonomy for urban sound research,''
\newblock in {\em 22st {ACM} International Conference on Multimedia
  ({ACM-MM'14})}, Orlando, FL, USA, Nov. 2014.

\bibitem{Mesaros2016_EUSIPCO}
Annamaria Mesaros, Toni Heittola, and Tuomas Virtanen,
\newblock ``{TUT} database for acoustic scene classification and sound event
  detection,''
\newblock in {\em 24th European Signal Processing Conference 2016 (EUSIPCO
  2016)}, Budapest, Hungary, 2016.

\bibitem{buckley2004retrieval}
C.~Buckley and E.~Voorhees,
\newblock ``Retrieval evaluation with incomplete information,''
\newblock in {\em Proceedings of the 27th annual international ACM SIGIR
  conference on Research and development in information retrieval}. ACM, 2004,
  pp. 25--32.

\end{thebibliography}

\end{document}